\documentclass{emulateapj}
\usepackage{apjfonts}
\usepackage{color}

\newcommand{\pivec}{\mbox{\boldmath $\pi$}}
\newcommand{\muvec}{\mbox{\boldmath $\mu$}}
\newcommand{\te}{t_{\rm E}}
\newcommand{\thetae}{\theta_{\rm E}}
\newcommand{\pie}{\pi_{\rm E}}
\newcommand{\pien}{\pi_{{\rm E},N}}
\newcommand{\piee}{\pi_{{\rm E},E}}
\newcommand{\dl}{D_{\rm L}}
\newcommand{\ds}{D_{\rm S}}

\definecolor{darkbrown}{RGB}{139,69,19}

\shorttitle{OGLE-2015-BLG-0196}
\shortauthors{HAN, UDALSKI, GOULD, ET AL.}
\begin{document}

\title{OGLE-2015-BLG-0196: Ground-based Gravitational Microlens Parallax Confirmed By Space-Based Observation}

\author{
C.~Han$^{1}$, A.~Udalski$^{2,11}$, A.~Gould$^{3,4,12}$, Wei~Zhu$^{3,12}$, 
\\
and\\
M.~K.~Szyma{\'n}ski$^{2}$, I.~Soszy{\'n}ski$^{2}$, J.~Skowron{2}, P.~Mr{\'o}z$^{2}$, 
R.~Poleski$^{2,3}$, P.~Pietrukowicz$^{2}$, S.~Koz{\l}owski$^{2}$, K.~Ulaczyk$^{2}$, 
M.~Pawlak$^{2}$\\
(The OGLE Collaboration),\\
J.~C.~Yee$^{5,13}$, C.~Beichman$^{6}$, S.~Calchi~Novati$^{6,7,8}$, S.~Carey$^{9}$, 
C.~Bryden$^{10}$, M.~Fausnaugh$^3$, B.~S.~Gaudi$^{3}$, Calen~B.~Henderson$^{10,14}$, 
Y.~Shvartzvald$^{10,14}$, B.~Wibking$^3$ \\ 
(The \textit{Spitzer} Microlensing Team),\\
}

\affil{$^{1}$  Department of Physics, Chungbuk National University, Cheongju 361-763, Republic of Korea}
\affil{$^{2}$  Warsaw University Observatory, Al. Ujazdowskie 4, 00-478 Warszawa, Poland}
\affil{$^{3}$  Department of Astronomy, Ohio State University, 140 W. 18th Ave., Columbus, OH 43210, USA}
\affil{$^{4}$  Max Planck Institute for Astronomy, K{\"o}nigstuhl 17, D-69117 Heidelberg, Germany}
\affil{$^{5}$  Harvard-Smithsonian Center for Astrophysics, 60 Garden St., Cambridge, MA 02138, USA}
\affil{$^{6}$  NASA Exoplanet Science Institute, MS 100-22, California Institute of Technology, Pasadena, CA 91125, USA}
\affil{$^{7}$  Dipartimento di Fisica "E.~R.~Caianiello", U\'niversit\'a di Salerno, Via Giovanni Paolo II, I-84084 Fisciano (SA), Italy}
\affil{$^{8}$ Infrared Processing and Analysis Center, NASA}
\affil{$^{9}$ Spitzer Science Center, MS 220-6, California Institute of Technology, Pasadena, CA, USA}
\affil{$^{10}$ Jet Propulsion Laboratory, California Institute of Technology, 4800 Oak Grove Drive, Pasadena, CA 91109, USA}
\footnotetext[11]{The OGLE Collaboration.}
\footnotetext[12]{The $Spitzer$ Microlensing Team.}
\footnotetext[13]{Sagan Fellow.}
\footnotetext[14]{NASA Postdoctoral Fellow.}

\begin{abstract}
In this paper, we present the analysis of the binary gravitational microlensing event 
OGLE-2015-BLG-0196. The event lasted for almost a year and the light curve exhibited 
significant deviations from the lensing model based on the rectilinear lens-source 
relative motion, enabling us to measure the microlens parallax. The ground-based microlens 
parallax is confirmed by the data obtained from space-based microlens observations using 
the {\it Spitzer} telescope.  By additionally measuring the angular Einstein radius from 
the analysis of the resolved caustic crossing, the physical parameters of the lens are 
determined up to the two-fold degeneracy: $u_0<0$ and $u_0>0$ solutions caused by the 
well-known ``ecliptic'' degeneracy.  It is found that the binary lens is composed of 
two M dwarf stars with similar masses $M_1=0.38\pm 0.04\ M_\odot$ ($0.50\pm 0.05\ M_\odot)$ 
and $M_2=0.38\pm 0.04\ M_\odot$ ($0.55\pm 0.06\ M_\odot$) and the distance to the lens 
is $D_{\rm L}=2.77\pm 0.23$ kpc ($3.30\pm 0.29$ kpc). Here the physical parameters out 
and in the parenthesis are for the $u_0<0$ and $u_0>0$ solutions, respectively.
\end{abstract}


\section{Introduction}

A microlens parallax represents the ratio of the relative lens-source parallax 
$\pi_{\rm rel}$ to the angular Einstein radius $\thetae$, i.e.
\begin{equation}
\pivec_{\rm E} ={\pi_{\rm rel} \over \thetae}{\muvec \over \mu}\qquad  
\pi_{\rm rel}={\rm au} \left( {1\over \dl}-{1\over \ds}\right),
\label{eq1}
\end{equation}
where $\muvec$ is the relative lens-source proper motion vector, $\dl$ and $\ds$  
denote the distances to the lens and source, respectively.  The microlensing parallax 
measurement is important because $\pie$ enables one to determine the mass and the 
distance to the lens through the relations \citep{Gould2000}
\begin{equation}
M={\thetae\over \kappa \pie};\qquad
\dl={{\rm au}\over \pie\thetae+\pi_{\rm S}},  
\label{eq2}
\end{equation}
where $\kappa=4G/(c^2 {\rm au})$ and $\pi_{\rm S}={\rm au}/\ds$.

For a small fraction of long time-scale events produced by nearby lenses, the microlens 
parallax can be measured in a single frame of the accelerating Earth. This so-called 
``annual microlens parallax'' is measured from the modulation in the lensing light 
curve caused by the  orbital motion of the Earth around the Sun \citep{Gould1992}.  
For most lensing events with known physical lens parameters, microlens parallaxes 
were measured through this channel.

The microlens parallax can also be measured if a lensing event is simultaneously 
observed from the ground-based observatory and from a satellite in a solar orbit 
\citep{Refsdal1966, Gould1994}. The measurement of this so-called ``space-based 
microlens parallax'' is possible because the projected Earth-satellite separation 
is comparable to the Einstein radius of typical Galactic microlensing events, i.e. 
$\sim {\cal O}$ (au), and thus the relative lens-source positions seen from the 
ground and from the satellite appear to be different.

The first space-based microlensing observations were conducted with the 
{\it Spitzer Space Telescope} for a lensing event occurred on a star in the Small 
Magellanic Cloud \citep[OGLE-2005-SMC-0001:][]{Dong2007} 41 years after S.~Refsdal 
first proposed the idea.  Space-based observations were also conducted with the 
use of the {\it Deep Impact} (or {\it EPOXI}) spacecraft for a planetary microlensing 
event \citep[MOA-2009-BLG-266:][]{Muraki2011}.  A space-based microlensing campaign 
making use of the {\it Spitzer} telescope to determine microlensing parallaxes has 
been operating since 2014.  The goal of the program is to determine the distribution 
of planets in the Galaxy by estimating the distances to individual lenses 
\citep{Calchi2015a}.  In addition, a space-based survey using the {\it Kepler} space 
telescope ({\it K2C9}), was conducted during the 2016 microlensing season.  The 
{\it K2} microlensing survey is expected to measure microlens parallaxes for 
$\gtrsim 127$ lensing events \citep{Henderson2016}.  The {\it Spitzer} microlensing 
campaign combined with ground-based survey and follow-up observations enabled the 
measurement of microlens parallaxes for various types of lenses, including single-mass 
objects \citep{Yee2015, Zhu2016}\footnote{For the lensing event OGLE-2015-BLG-0763 
\citep{Zhu2016}, the {\it Spitzer} observation enabled to uniquely determine the mass 
of an isolated star by measuring both $\pi_{\rm E}$ and $\theta_{\rm E}$. }, planetary 
systems \citep{Udalski2015b, Street2016}, and binary systems 
\citep{Zhu2015, Shvartzvald2015, Shvartzvald2016, Bozza2016, Han2016a}.  For all of 
these events, the physical parameters of the lenses were constrained using space-based 
microlens parallaxes.  However, except for OGLE-2014-BLG-0124 \citep{Udalski2015b}, 
the measured microlens parallaxes have not been confirmed by the annual parallax 
measurements from ground-based observations because event time scales are not 
sufficiently long enough to allow measurement of the annual parallaxes.

In this work, we report the results from the analysis of the binary-lens microlensing 
event OGLE-2015-BLG-0196 that was observed simultaneously from the ground and from 
the {\it Spitzer} telescope.  The ground-based light curve shows significant deviations 
from the standard model based on the rectilinear relative lens-source motion, enabling 
us to measure the microlens parallax.  The ground-based microlens parallax estimate is 
confirmed by the {\it Spitzer} observations.

\section{Observation}

OGLE-2015-BLG-0196 involved a star located toward the Galactic Bulge field, in the 
field BLG660.12 of the OGLE-IV survey.  The equatorial coordinates of the event are 
$(\alpha,\delta)_{\rm J2000}=(17^\circ 45'58''\hskip-2pt.3, -32^{\rm h}57^{\rm m}24^{\rm s}\hskip-2pt.4)$,
which correspond to the Galactic coordinates
$(l,b)=(356^\circ\hskip-2pt.61, -2^\circ\hskip-2pt.16)$. The lensing-induced 
brightening of the star was discovered on 2015 February 26 by the Early Warning System 
\citep[EWS:][]{Udalski2003} of the fourth phase of the Optical Gravitational Lensing 
Experiment \citep[OGLE-IV:][]{Udalski2015a} microlensing survey.  The OGLE survey uses 
the 1.3m Warsaw telescope located at the Las Campanas Observatory in Chile.

\begin{figure}
\includegraphics[width=\columnwidth]{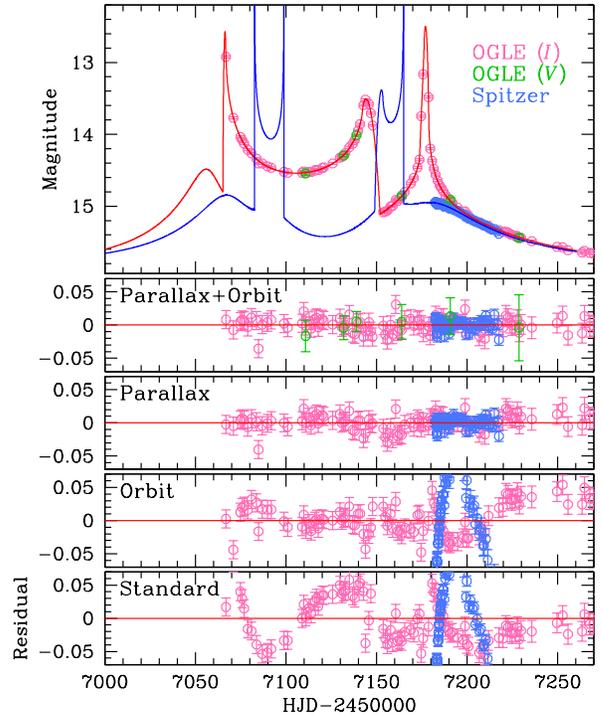}
\caption{Light curve of OGLE-2015-BLG-0196. The pink and light-blue dots in the upper panels 
show the data obtained from the ground-based observatory and Spitzer telescope, respectively. 
The lower four panels show the residuals from the 4 tested models. The solid curve superposed 
on the data points is the best-fit model (``parallax+orbit'' model). }
\label{fig:one}
\end{figure}

When the event was discovered, the light curve already deviated from the symmetric shape 
of a single-mass event.  As the event progressed, the light curve exhibited a ``U''-shape 
feature, which is a characteristic feature occurring when a source star passes the inner 
region of a binary-lens caustic. Since binary caustics form a closed curve, a caustic 
exit was expected and it actually happened on ${\rm HJD}'={\rm HJD}-2450000\sim 7143$.  
From the preliminary modeling of the lensing light curve conducted after the caustic 
exit, it was anticipated that there would be another caustic-crossing feature at 
${\rm HJD}'\sim 7177$. The caustic feature occurred as predicted by the model.  
Subsequently, the light curve gradually returned to baseline.

\begin{figure}
\includegraphics[width=\columnwidth]{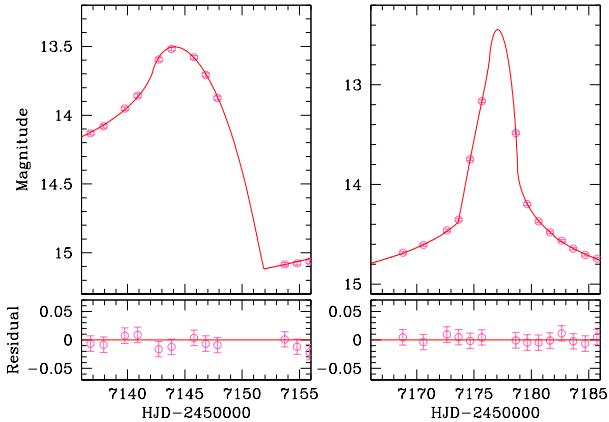}
\caption{Zoomed view of the light light curve around the second and third peaks at 
${\rm HJD}'={\rm HJD}-2450000 \sim 7143$ and 7177.  }
\label{fig:two}
\end{figure}

In Figure~\ref{fig:one}, we present the light curve of OGLE-2015-BLG-0196, where the pink 
dots are the data obtained from the ground-based observations. The light curve is composed 
of 3 peaks that occurred at ${\rm HJD}'\sim 7065$, 7143, and 7177. The first two peaks 
correspond to a caustic entrance and exit.  The wide time gap of $\sim 78$ days between 
the caustic-crossing features indicates that the features resulted from the source crossing 
a large caustic.  On the other hand, the third peak does not show a characteristic U-shape 
feature, suggesting the feature resulted from the source crossing over the cusp of the 
caustic.  We note that the second and third caustic-crossing features were well resolved 
by ground-based observations.  See the zoomed view of the resolved caustic-crossing features 
presented in Figure~\ref{fig:two}.  The first caustic-crossing feature was not resolved 
because it occurred before the start of the 2015 bulge season. Another factor to be noted 
is the slow progress of the event. The event occurred before the beginning of the 2015 
bulge season and proceeded throughout the whole bulge season.

The event was selected as a target for {\it Spitzer} observations because of the chance 
to measure the parallax effect between {\it Spitzer} and the Earth, which were separated 
by a projected separation of $D_{\perp} \sim 1.4\,$ au. It is particularly interesting 
to measure this effect for this binary since the caustic crossings are well resolved, 
meaning that the angular Einstein radius, and thereby the lens mass, could be determined 
from the measured $\theta_{\rm E}$ and $\pi_{\rm E}$ following Eq.~(\ref{eq2}). This 
binary was selected subjectively because it did not meet the objective binary criteria as 
described in \citet{Yee2015}.  The {\it Spitzer} observations of the event were conducted 
for $\sim 35$ days  from ${\rm HJD}'=7182.4$ to ${\rm HJD}'=7217.5$. The observation 
cadence varies in the range $\sim 1/9$ -- 1 day and the total number of data points is 65. 
The {\it Spitzer} data are presented in the upper panel of Figure~\ref{fig:one} (blue dots).

Reduction of the ground-based data was done using the Difference Imaging Analysis pipeline 
\citep{Udalski2003} of the OGLE survey.  The {\it Spitzer} data were reduced using the 
algorithm specialized for {\it Spitzer} photometry in crowded fields \citep{Calchi2015b}.  
For the individual data sets, we readjust error bars by
\begin{equation}
\sigma=k(\sigma_0^2+\sigma_{\rm min}^2)^{1/2},
\label{eq3}
\end{equation}
where $\sigma_0$ represents the error bar estimated from the automatized pipeline,
$\sigma_{\rm min}$ is the factor used to make the error bars be consistent with 
the scatter of data points, and the other factor $k$ is used to make $\chi^2/{\rm dof}=1$. 
The adopted values of the scaling factor $k$ and the minimum error $\sigma_{\rm min}$ are
$k=2.33$ and 0.40 
and
$\sigma_{\rm min}=0.005$ mag and 0.020 mag,
for the OGLE and {\it Spitzer} data sets, respectively.\footnote{The data sets used for the 
analysis are posted at the web site http://astroph.chungbuk.ac.kr/$\sim$cheongho/OB150196/data.html
for the independent verification of the results.}

\section{Analysis}

The number of parameters needed to model a binary event light curve in the simplest case of 
a rectilinear relative lens-source motion is 7 (principal parameters) plus 2 flux parameters 
for each data set.  The principal parameters include the epoch of the closest lens-source 
approach, $t_0$, the lens-source separation at $t_0$, $u_0$, the Einstein time scale, $\te$, 
the separation $s$ and the mass ratio $q$ between the binary lens components, the angle between 
the source trajectory and the binary axis, $\alpha$, and the normalized source radius $\rho$.  
We note that lengths of the parameters $u_0$, $s$, $\rho$ are normalized to the angular Einstein 
radius $\thetae$ and the Einstein time scale $\te$ represents the time interval for the source 
to traverse $\thetae$.  For the reference position of the lens, we use the barycenter of the 
binary lens.  The flux parameters $F_{\rm S}$ and $F_{\rm B}$ represent the flux from the 
source and blend, respectively.

We start modeling of the event light curve with the 7 principal parameters (``standard model''). 
Modeling is performed in several steps.  In the first step, a grid search is conducted for the 
parameters $s$, $q$, and $\alpha$ for which the lensing magnification is sensitive to small 
changes of the the parameters.  The other parameters are optimized using a downhill approach, 
where we use the Markov Chain Monte Carlo (MCMC) method.  In the second step, we locate local 
minima in the $\chi^2$ map of the parameters in order to check the existence of possible 
degenerate solutions which result in similar light curves despite the combinations of widely 
different parameters.  In the last step, a global solution is identified from the comparison 
of the local solutions.

In computing lensing magnifications, we consider finite-source effects.  Finite magnifications 
are computed by using both numerical and semi-analytic methods. In the region very close to 
caustics, we use the numerical ray-shooting method \citep{Schneider1986}.  In the region around 
the caustic, we use the semi-analytic hexadecapole approximation \citep{Pejcha2009, Gould2008}.  
We also consider the limb-darkening effect of the source star.  For this, the surface brightness 
variation is parameterized as $S_\lambda \propto 1-\Gamma_\lambda(1-3\cos\phi/2)$, where $\lambda$ 
denotes the filter used for observation and $\phi$ represents the angle between the normal to 
the source star's surface and the line of sight toward the center of the source star.  The 
adopted value of the limb-darkening coefficient is $\Gamma_I=0.62$, which is chosen from the 
catalog of \citet{Claret2000} based on the stellar type of the source star.  The source type 
is determined from the de-reddened color and brightness for which we discuss about the procedure 
of the determinations in section 4.

From the standard modeling, we find a unique solution that describes the three main 
caustic-crossing features of the ground-based light curve.  From the solution, it is found 
that the lens is a binary object composed of similar-mass components and a projected separation 
slightly greater than $\thetae$. However, the solution leaves a significant residual from the 
model as presented in the bottom panel of Figure~\ref{fig:one}. The residual persists throughout 
the event, indicating that one should consider higher-order effects that cause long-lasting 
deviations.  To be also noted is that the standard model provides a poor fit to the 
{\it Spitzer} data.

It is known that the orbital motion of a binary lens can cause long-term deviations in 
lensing light curves \citep{Albrow2000, Shin2011, Park2013}.  Consideration of the orbital 
effect requires to include two additional parameters $ds/dt$ and $d\alpha/dt$, where $ds/dt$ 
denotes the rate of the binary separation change and $d\alpha/dt$ represents the change rate 
of the source trajectory angle.  From the modeling considering the orbital effect (``orbit model''), 
we find that the observed data still leave a substantial residual from the model, indicating 
that the orbital effect is not the main cause of the deviation.  In the third residual panel 
of Figure~\ref{fig:one}, we present the residual from the orbit model.

Another higher-order effect known to cause long-term deviations is the annual parallax 
effect.  We check the possibility of the parallax effect by conducting  another modeling 
considering the parallax effect (``parallax model'').  This requires modeling with two 
additional parameters $\pien$ and $\piee$, which are the the components of $\pivec_{\rm E}$ 
projected onto the sky along the north and east equatorial coordinates, respectively.

\begin{deluxetable}{lcrr}
\tablecaption{Comparison of Models\label{table:one}}
\tablewidth{0pt}
\tablehead{
\multicolumn{2}{c}{Model}        &
\multicolumn{2}{c}{$\chi^2$}     \\
\multicolumn{1}{c}{}        &
\multicolumn{1}{c}{}        &
\multicolumn{1}{c}{OGLE+{\it Spitzer}}        &
\multicolumn{1}{c}{OGLE only}       
}
\startdata
 Standard               &             & 3123.7  &  1027.0   \\
 Orbit                  &             & 2489.2  &   609.7  \\ 
 Parallax               &  ($u_0<0$)  &  596.1  &   549.6  \\
                        &  ($u_0>0$)  &  625.7  &   551.2  \\
{\bf Parallax + Orbit}  &  ($u_0<0$)  &  571.7  &   524.0  \\
                        &  ($u_0>0$)  &  572.8  &   525.1    
\enddata                                                                                              
\end{deluxetable}                                                                                    

Although parallax effects on the light curves obtained from ground-based and space-based 
observations manifest in different ways, the microlens parallax values measured through 
the different channels of the annual and the space-based microlens parallax observations should 
be the same. This implies that if the parallax effect detected in the light curve obtained from 
the ground-based observation is real, the effect should also be able to explain the light curve 
obtained from the {\it Spitzer} observation. Therefore, we conduct two sets of modeling where 
the first modeling is based on the ground-based data, while the second modeling is based on 
the combined data sets from the ground-based and from the {\it Spitzer} observations.  From 
these modelings, we find that the parallax effect can explain both the deviations of the 
ground-based and the {\it Spitzer} data, as shown in the second residual panel of
Figure~\ref{fig:one}.  This indicates that the major cause of the deviation from the 
standard model is the parallax effect.

In the modeling considering parallax effects, we check the existence of degenerate solutions. 
It is a well-known fact that analyzing light curves of single-mass lensing events obtained from 
both space- and ground-based observations yields four sets of degenerate solutions \citep{Gould1994}, 
which are often denoted by $(+,+)$, $(-,-)$, $(+,-)$, and $(-,+)$, where the former and latter signs 
in each parenthesis represent the signs of the lens-source impact parameters as seen from Earth and 
from the satellite, respectively. In the case of binary-lensing, the four-fold parallax degeneracy 
collapses into a two-fold degeneracy for a general case of binary-lens 
events because the degeneracy between the pair of $(+,+)$ and $(+,-)$ [or $(-,-)$ and $(-,+)$] 
solutions are generally resolved due to the lack of lensing magnification symmetry compared to the 
single lens case although the remaining degeneracy, i.e.\ $(+,+)$ and $(-,-)$ solutions, may persist. 
However, \citet{Zhu2015} pointed that the four-fold degeneracy can persist in some special cases of 
the lens-source geometry. We, therefore, check the degeneracy by conducting a grid search in the 
$\pien$-$\piee$ parameter space. From this, we find that the light curve of OGLE-2015-BLG-0196 does 
not suffer from the degeneracy between $(\pm,\pm)$ and $(\pm,\mp)$ solutions and only the degeneracy 
between the $(+,+)$ and $(-,-)$ solutions persists.  This degeneracy between the $(+,+)$ and $(-,-)$  
solutions, which is referred to as the ``ecliptic degeneracy'' \citep{Skowron2011}, is known to exist 
for general binary-lens events.  This degeneracy is caused because the two source trajectories with 
$u_0$ and $-u_0$ are in mirror symmetry with respect to the binary axis.  For this reason, the 
degenerate $(+,+)$ and $(-,-)$ solutions are often denoted by $u_0 > 0$ and $u_0 < 0$ solutions, 
respectively.  We note that the lensing parameters of the degenerate solutions caused by the 
ecliptic degeneracy are in the relation 
$(u_0,\alpha,\pien,d\alpha/dt) \leftrightarrow -(u_0,\alpha,\pien,d\alpha/dt)$ \citep{Skowron2011}.

\begin{figure}
\includegraphics[width=\columnwidth]{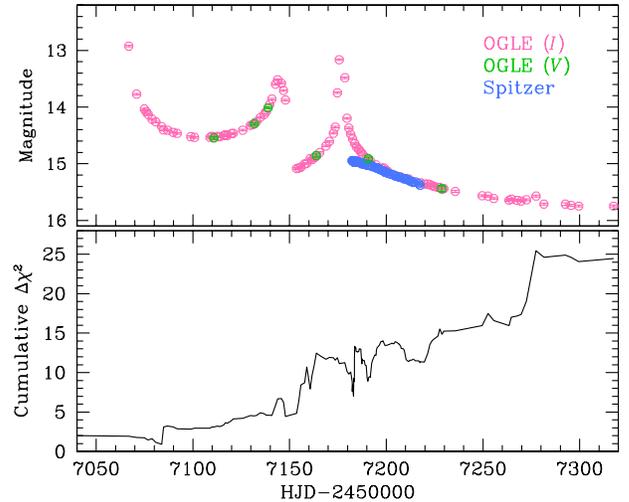}
\caption{ Cumulative distribution of $\Delta\chi^2$ between the "parallax only" model and the 
"parallax + orbit" model. The lensing light curve in the upper panel is presented to show
$\chi^2$ improvement with the progress of the event.
}
\label{fig:three}
\end{figure}

\begin{deluxetable*}{lrrrr}
\tablecaption{Best-fit Lensing Parameters\label{table:two}}
\tablewidth{0pt}
\tablehead{
\multicolumn{1}{c}{Parameter}        &
\multicolumn{2}{c}{OGLE+{\it Spitzer}}        &
\multicolumn{2}{c}{OGLE only}     \\
\multicolumn{1}{c}{}        &
\multicolumn{1}{c}{$u_0<0$}          &
\multicolumn{1}{c}{$u_0>0$}          &
\multicolumn{1}{c}{$u_0<0$}          &
\multicolumn{1}{c}{$u_0>0$}          
}
\startdata
  $\chi^2$                     &         571.7         &          572.8        &        524.0          &         525.1          \\  

  $t_0$ (HJD-2450000)          & 7115.561 $\pm$ 0.916  & 7119.410 $\pm$  0.954 &  7115.506 $\pm$ 0.489 &   7121.000 $\pm$ 0.968 \\  
  $u_0$                        &   -0.041 $\pm$ 0.004  &    0.044 $\pm$  0.003 & -0.037    $\pm$ 0.004 &   0.048    $\pm$ 0.004 \\  
  $t_{\rm E}$ (days)           &   96.7   $\pm$ 0.6    &   92.9   $\pm$  0.6   &  96.7     $\pm$ 0.6   &   92.9     $\pm$ 0.6   \\  
  $s$                          &    1.55  $\pm$ 0.01   &    1.61  $\pm$  0.01  &  1.55     $\pm$ 0.01  &   1.63     $\pm$ 0.01  \\  
  $q$                          &    1.01  $\pm$ 0.02   &    1.10  $\pm$  0.03  &  1.01     $\pm$ 0.02  &   1.15     $\pm$ 0.03  \\  
  $\alpha$ (rad)               &    0.292 $\pm$ 0.013  &   -0.217 $\pm$  0.011 &  0.304    $\pm$ 0.013 &  -0.248    $\pm$ 0.011 \\  
  $\rho$ ($10^{-3}$)           &    8.31  $\pm$ 0.18   &    7.80  $\pm$  0.13  &  8.57     $\pm$ 0.21  &   7.79     $\pm$ 0.19  \\  
  $\pi_{{\rm E},N}$            &    0.171 $\pm$ 0.011  &   -0.116 $\pm$  0.011 &  0.198    $\pm$ 0.016 &  -0.164    $\pm$ 0.018 \\  
  $\pi_{{\rm E},E}$            &    0.105 $\pm$ 0.007  &    0.093 $\pm$  0.006 &  0.100    $\pm$ 0.009 &   0.097    $\pm$ 0.014 \\  
  $ds/dt$     (yr$^{-1}$)      &    0.27  $\pm$ 0.06   &    0.01  $\pm$  0.05  &  0.30     $\pm$ 0.06  &  -0.11     $\pm$ 0.08  \\  
  $d\alpha/dt$ (rad yr$^{-1}$) &   -0.20  $\pm$ 0.05   &   -0.29  $\pm$  0.03  & -0.31     $\pm$ 0.06  &  -0.15     $\pm$ 0.06  \\  
  $f_{{\rm S},I}$              &    7.27  $\pm$ 0.03   &    7.02  $\pm$  0.05  &  7.27     $\pm$ 0.03  &   7.02     $\pm$  0.05 \\  
  $f_{{\rm B},I}$              &    0.11  $\pm$ 0.04   &    0.36  $\pm$  0.05  &  0.11     $\pm$ 0.04  &   0.36     $\pm$  0.05 \\
  $f_{{\rm S},V}$              &    0.16  $\pm$ 0.01   &    0.16  $\pm$  0.01  &  0.16     $\pm$ 0.01  &   0.16     $\pm$  0.01   
\enddata                                                                                              
\tablecomments{The source and blending fluxes are normalized so that $f=1$ for an $I=18$ star.}
\end{deluxetable*}                                                                                    

\citet{Han2016b} pointed out that space-based microlens parallax observations can be useful not 
only for the microlens parallax measurement but also for the measurement of the orbital parameters. 
This is possible because the difference between the light curves seen from the ground and from 
a solar-orbit satellite produces a large parallax effect. At the same time, the features of the 
binary light curve as seen from the ground give precise timings for the caustic crossings. In 
the absence of space observations, these features give a measurement of the combination of 
parallax and orbital motion of the binary (which are partially degenerate \citep{Batista2011, 
Skowron2011}). However, the microlens parallax is already mostly determined from the space-based 
parallax effect, so the information from the caustic crossing timing goes almost entirely into 
measuring the orbital motion.  We, therefore, conduct an additional modeling, where both the 
orbital and the parallax effects are taken into account (``parallax + orbit'' model).  We find 
that the additional consideration of the orbital effect further improves the fit by $\Delta\chi^2=24.2$.  
Due to the small $\chi^2$ difference between the parallax only model and the parallax+orbit model, 
the improvement of the fit is is not immediately clear in the residuals.  In Figure~\ref{fig:three}, 
we present the cumulative $\Delta\chi^2$ distribution to better show the improvement of the fit by 
the orbital effect.  One finds that the fit improvement occurs throughout the event.

\begin{figure}
\includegraphics[width=\columnwidth]{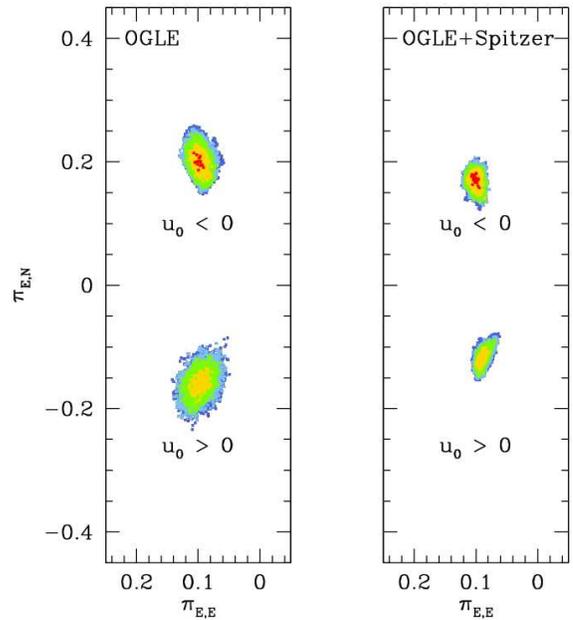}
\caption{
Distribution of the microlens parallax parameters $\pi_{{\rm E},N}$ and $\pi_{{\rm E},E}$.  
The distributions in the left and right panels are obtained based on the ground-based OGLE 
data and the combined OGLE+{\it Spitzer} data, respectively.  The color coding indicates 
points on the Markov Chain within 1$\sigma$ (red), 2$\sigma$ (yellow), 3$\sigma$ (green), 
4$\sigma$ (cyan), and 5$\sigma$ (blue) of the best fit.}
\label{fig:four}
\end{figure}

\begin{figure}
\includegraphics[width=\columnwidth]{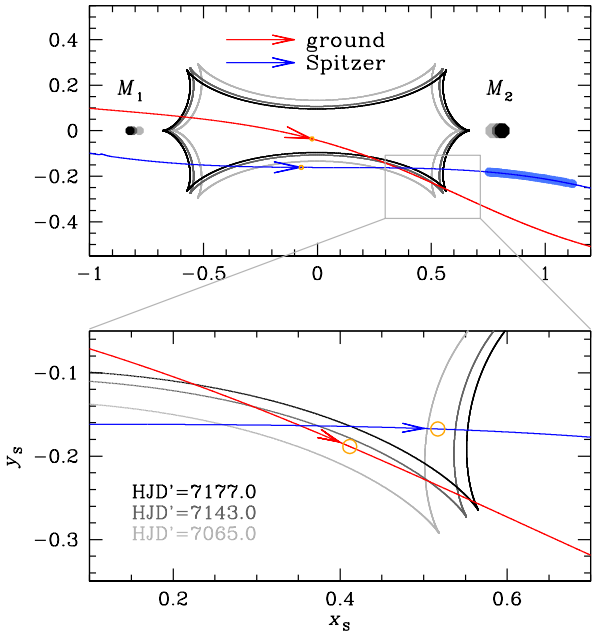}
\caption{Geometry of the lens system. The closed curve with six cusps is the caustic and the 
blue and red curves with arrows represent the source trajectories seen from the ground and from 
the {\it Spitzer} telescope, respectively. The caustic varies in time due to the orbital motion 
of the binary lens and we present 3 caustics corresponding to 3 different times marked in the 
lower panel. The lower panel shows the zoom of the region enclosed by a small box in the upper 
panel. The small orange circle at the tip of the arrow on the source trajectory represents the 
source size. The small filled dots marked by $M_1$ and $M_2$ are the binary-lens components.
We note that $M_1$ is lighter in mass than $M_2$ because $M_1$ is defined as the lens component
lying closer to the source trajectory.  The points on the trajectory seen from the {\it Spitzer} 
telescope represent the source positions during the {\it Spitzer} observation.  The coordinates 
are centered at the barycenter of the binary lens and all lengths are normalized to the angular 
Einstein radius corresponding to the total mass of the binary lens.}
\label{fig:five}
\end{figure}

In Table~\ref{table:one}, we present the $\chi^2$ values of the tested models in order to compare 
the goodness of the individual fits.  For each model, we present both $\chi^2$ values determined 
based on the ground-based OGLE data and the combined OGLE+{\it Spitzer} data.

Figure~\ref{fig:four} shows the distributions of the microlens parallax parameters 
$\pi_{{\rm E},N}$ and $\pi_{{\rm E},E}$ that are determined based on two different sets of data: 
one based on the ground-based OGLE data and the other based on the combined OGLE+{\it Spitzer} 
data.  From the comparison of the distributions, one finds that the parallax parameters 
determined based on the two data sets match very well, indicating that the ground-based 
microlensing parallax is confirmed by the space-based observation.  One also finds that the 
uncertainties of the parallax parameters based on the combined data is substantially smaller 
than the uncertainties based on only the ground-based data.  This indicates that the 
{\it Spitzer} data add an important constraint on the parallax measurement despite the 
short coverage of the event.

In Table~\ref{table:two}, we list the best-fit lensing parameters of the ``parallax + orbit 
model'' along with $\chi^2$ values.  For comparison, we also present the parameters obtained 
based on the ground-based data.  We find that the degeneracy between the $u_0 < 0$ and $u_0 > 0$ 
solution is very severe ($\Delta\chi^2=1.1$) and thus present both solutions. The estimated values 
of the normalized separation and the mass ratio between the binary lens components are 
$(s,q)=(1.55 \pm 0.01, 1.01 \pm 0.02)$ for the $u_0 < 0$ solution and 
$(s,q)=(1.61 \pm 0.01, 1.10 \pm 0.03)$ for the $u_0 > 0$ solution, indicating that the binary 
components have similar masses and the projected separation is $\sim 1.6$ times greater than 
the Einstein radius.  We note that $q > 1.0$ implies that the the lens component with the 
smaller separation from the source trajectory, $M_1$, is lighter in mass than the other lens 
component, $M_2$.

Figure~\ref{fig:five} shows the lens system geometry (for the parallax + orbit model with $u_0<0$), 
where the trajectories of the source star with respect to the lens and the caustic are presented.
We note that two source trajectories are presented: one seen from the ground (red curve) and the 
other seen from the {\it Spitzer} telescope (blue curve).  Since the binary separation is not 
much different from the Einstein radius, the caustic is composed of a single big closed curve 
(resonant caustic) and it is elongated along the binary axis because $s > 1$.  We note that 
the caustic changes in time because the orbital motion of the binary lens causes the separation 
and the orientation of the binary lens to vary in time.

The ground-based source trajectory entered the upper left part of the caustic, diagonally passed 
the caustic, and then exited the caustic. Due to the concavity of the caustic curve, the source 
reentered the tip of the caustic, and then exited. It is found the two caustic-crossing spikes 
at ${\rm HJD}'=7065$ and 7143 in the ground-based light curve were produced by the first set of 
the caustic entrance and exit, while the caustic feature at ${\rm HJD}'=7177$ was produced by the 
second set of the caustic entrance and exit. The reason why the caustic-crossing feature at 
${\rm HJD}'=7177$ does not show a characteristic U-shape feature is that the width of the caustic 
tip is smaller than the source star and thereby the U-shape feature in the light curve is smeared 
out by finite-source effects. See the lower panel of Figure~\ref{fig:five} where we present the 
zoomed view of the caustic tip.

The source seen from the {\it Spitzer} telescope took a different trajectory from the one seen 
from the ground.  The source moved almost in parallel with the binary axis during which the 
caustic experienced two sets of caustic entrance and exit. The part of the light curve observed 
by the {\it Spitzer} telescope (marked by blue dots on the source trajectory in the upper panel 
of Figure~\ref{fig:five}) corresponds to the declining part after the second caustic exit.

\begin{figure}
\includegraphics[width=\columnwidth]{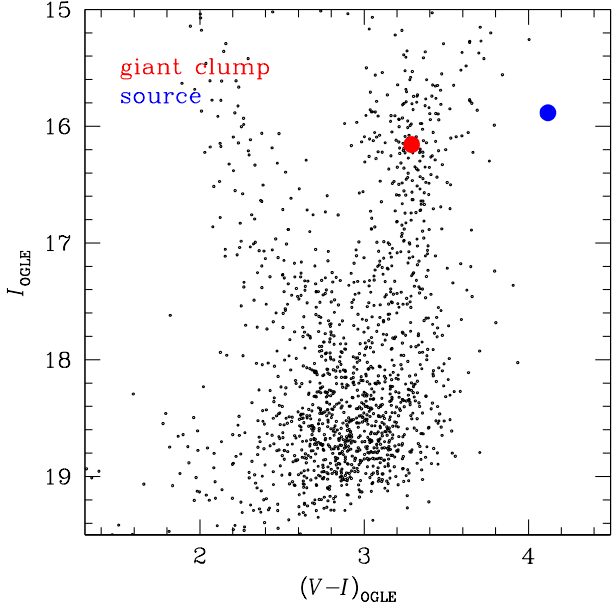}
\caption{Location of the source star in the color-magnitude diagram of stars in the neighboring 
region around the source star. The blue and red dots represent the source star and 
the centroid of giant clump, respectively.}
\label{fig:six}
\end{figure}

\begin{deluxetable*}{lrrrr}
\tablecaption{Physical Parameters\label{table:three}}
\tablewidth{0pt}
\tablehead{
\multicolumn{1}{c}{Parameter}        &
\multicolumn{2}{c}{OGLE+{\it Spitzer}}        &
\multicolumn{2}{c}{OGLE only}     \\
\multicolumn{1}{c}{}                 &
\multicolumn{1}{c}{$u_0<0$}          &
\multicolumn{1}{c}{$u_0>0$}          &
\multicolumn{1}{c}{$u_0<0$}          &
\multicolumn{1}{c}{$u_0>0$}           
}
\startdata
  $M_1$      ($M_\odot$)                    &  0.38 $\pm$ 0.04  &  0.50 $\pm$ 0.05  &  0.32 $\pm$ 0.04  &  0.39 $\pm$ 0.05 \\    
  $M_2$      ($M_\odot$)                    &  0.38 $\pm$ 0.04  &  0.55 $\pm$ 0.06  &  0.32 $\pm$ 0.04  &  0.44 $\pm$ 0.06 \\      
  Distance to lens (kpc)                    &  2.77 $\pm$ 0.23  &  3.30 $\pm$ 0.29  &  2.66 $\pm$ 0.23  &  2.78 $\pm$ 0.29 \\      
  projected separation  (AU)                &  5.30 $\pm$ 0.43  &  6.77 $\pm$ 0.59  &  4.83 $\pm$ 0.44  &  5.87 $\pm$ 0.59 \\    
  Geocentric proper motion (mas yr$^{-1}$)  &  4.66 $\pm$ 0.37  &  5.00 $\pm$ 0.40  &  4.43 $\pm$ 0.38  &  5.08 $\pm$ 0.41 \\       
  KE/PE                                     &  0.17             &  0.32             &  0.30             &  0.09                                          
\enddata                                                                                              
\end{deluxetable*} 

\section{Lens Parameters}

For the unique determination of the lens mass and distance, one additionally needs the angular 
Einstein radius in addition to the microlens parallax.  The angular Einstein radius is determined 
from the normalized source radius $\rho$ and the 
angular source radius $\theta_*$ by
\begin{equation}
\thetae={\theta_* \over \rho}.
\label{eq4}
\end{equation}
The value of $\rho$ is determined by analyzing the 
resolved caustic crossings that are affected by finite-source effects.

We estimate the angular source radius from the stellar type determined based on the de-reddened 
color $(V-I)_0$ and brightness $I_0$.  For this, we first measure the instrumental $I$-band 
magnitude from the flux parameters $F_{{\rm S},I}$ and $F_{{\rm B},I}$ and the instrumental 
color $V-I$ from the source flux $F_{{\rm S},I}$ and $F_{{\rm S},V}$ determined from the modeling 
based on the $I$- and $V$-band OGLE data.  In Figure~\ref{fig:six}, we mark the source position 
$(V-I,I)=(4.16,15.85)$ in the instrumental color-magnitude diagram of the field around the source 
star of OGLE-2015-BLG-0196.  We calibrate the color and brightness of the source star using the 
giant clump (GC) centroid in the color-magnitude diagram \citep{Yoo2004}.  The centroid of GC, 
marked by a red dot in Figure~\ref{fig:six}, can be used for calibration because (1) the de-reddened 
color $(V-I)_{0,{\rm GC}}=1.06$ \citep{Bensby2011} and the magnitude $I_{0,{\rm GC}}= 14.6$ 
\citep{Nataf2013} are known and (2) the source and GC stars are located in the bulge and thus 
experience a similar amount of extinction.  The source distance is estimated using the relation 
$D_{\rm S}=D_{\rm GC}/(\cos l + \sin l/\tan\phi)$ \citep{Nataf2013}, where $D_{\rm GC}=8160$ pc 
is the galactocentric distance and $\phi=40^\circ$ is the angle between the bulge's major axis 
and the line of sight.  From the difference in color and magnitude between the source star and 
the GC centroid, we estimate $(V-I,I)_0=(1.87, 14.3)$, indicating that the source is a very 
red M-type giant.  We covert $V-I$ into $V-K$ using the relation provided by \citet{Bessell1988}, 
and then derive $\theta_*$ from the relation between the $V-K$ and surface brightness 
\citep{Kervella2004}.  It is estimated that the source star has an angular radius of 
\begin{equation}
\theta_* = 10.25 \pm 0.83\ \mu{\rm as}.      
\label{eq5}
\end{equation}
From the measured $\theta_*$ and $\rho$, it is estimated that the angular Einstein radius 
of the lens system is 
\begin{equation}
\thetae = 1.23 \pm 0.10\ {\rm mas} \ \  (1.27 \pm 0.10\ {\rm mas}),
\label{eq6}
\end{equation}
where the values in and out of the parenthesis are the values for the $u_0<0$ and $u_0>0$ solution, 
respectively.

By measuring both $\thetae$ and $\pie$, the masses of the lens components are determined as
\begin{equation}
M_1 = 0.38 \pm 0.04\ M_\odot\ (0.50 \pm 0.05\ M_\odot)
\label{eq7}
\end{equation}
for the lens component located closer to the source trajectory and 
\begin{equation}
M_2 = 0.38 \pm 0.04\ M_\odot\ (0.55 \pm 0.06\ M_\odot)
\label{eq8}
\end{equation}
for the other lens component.
The distance to the lens is 
\begin{equation}
\dl = 2.77 \pm 0.23 \ {\rm kpc}\  (3.30 \pm 0.29\ {\rm kpc}).
\label{eq9}
\end{equation}
The estimated mass and distance indicate that the binary lens is composed of roughly 
equal-mass M dwarf stars and located in the disk of the Galaxy. 
The binary components are separated in projection by
\begin{equation}
r_\perp = 5.30 \pm 0.43\ {\rm au}\   (6.77 \pm 0.59\ {\rm au}). 
\label{eq10}
\end{equation}
In Table~\ref{table:three}, we summarize the physical lens parameters.  We note that the 
notation ``KE/PE'' represents the ratio of the transverse kinetic to potential energy 
that is computed by
\begin{equation}
\left( {{\rm KE}\over{\rm PE}}\right)_\perp =
{(r_\perp/{\rm au})^3 \over 8 \pi^2(M/M_\odot) }
\left[ \left({ 1\over s}{ds\over dt} \right)^2 + \left( {d\alpha\over dt}\right)^2\right],
\label{eq11}
\end{equation}
where $M=M_1+M_2$.  The ratio should be less than the three-dimensional kinetic to potential 
energy ratio, KE/PE, and should be less than unity for the system to be bound, 
i.e.~$({\rm KE}/{\rm PE})_\perp \leq {\rm KE}/{\rm PE} < 1$.  The determined 
$({\rm KE}/{\rm PE})_\perp$ satisfies this requirement.

\section{Conclusion}

We analyzed the binary-lensing event OGLE-2015-BLG-0196 that was observed both from 
the ground and from the {\it Spitzer Space Telescope}.  The light curve obtained 
from ground-based observations exhibited significant deviations from the lensing 
model based on the rectilinear relative lens-source motion motion and analysis of 
the deviation allowed us to measure the microlens parallax.  The measured microlens 
parallax was confirmed by the data obtained from space-based observations up to the 
two-fold degeneracy caused by the well-known ecliptic degeneracy.  This event is the 
first case where the ground-based microlens parallax was firmly confirmed with 
space-based observations.  By additionally measuring the angular Einstein radius 
from the analysis of caustic crossings of the light curve, we determined the mass 
and distance to the lens.  It was found that the lens is a binary composed of roughly 
equal-mass M dwarf stars located in the Galactic disk.

\begin{acknowledgments}
Work by C.~Han was supported by the Creative Research Initiative Program (2009-0081561) of 
National Research Foundation of Korea.  
The OGLE project has received funding from the National Science Centre, Poland, grant 
MAESTRO 2014/14/A/ST9/00121 to AU.  OGLE Team thanks Profs.\ M.~Kubiak, G.~Pietrzy{\'n}ski, 
and {\L}.~Wyrzykowski$^{2}$, former members of the OGLE team, for their contribution to 
the collection of the OGLE photometric data over the past years.
Work by AG was supported by JPL grant 1500811.
WZ acknowledges the support from NSF grant AST-1516842. 
Work by J.C.Y. was performed under contract with
the California Institute of Technology (Caltech)/Jet Propulsion
Laboratory (JPL) funded by NASA through the Sagan
Fellowship Program executed by the NASA Exoplanet Science
Institute.
Work by CBH and YS was supported by an appointment to the NASA Postdoctoral 
Program at the Jet Propulsion Laboratory, administered by Universities Space 
Research Association through a contract with NASA.
We acknowledge the high-speed internet service (KREONET)
provided by Korea Institute of Science and Technology Information (KISTI).

\end{acknowledgments}

\end{document}